# GENERAL NEUTRINO MASS MATRIX PATTERNS
# AND ITS UNDERLYING FAMILY SYMMETRIES

A. Damanik[a], M. Satriawan[b], P. Anggraita[c], and Muslim[b]

[a] Physics Department Faculty of Mathematics and Natural Sciences
Sanata Dharma University, Jogjakarta
[b] Physics Department Faculty of Mathematics and Natural Sciences
Gadjah Mada University, Jogjakarta
[c] National Nuclear Energy Agency (BATAN), Jakarta

**Abstract**
Based on current experimental results, such as neutrino oscillations, and the neutrinoless double beta decays (i.e. data from Super-Kamiokande, KamLAND, SNO, etc.), the neutrino mixing matrix can be adequately determined. Though there are still certain parameters that have large possibility limits, but based on the current experimental results, it is possible to construct a general form of neutrino mass matrix.
Starting from this general form of the neutrino mass matrix, we put certain conditions in the context of the seesaw mechanism model to determine the possible pattern of the neutrino mass matrix that has texture zero. From the obtained neutrino mass matrix pattern, there are three class of pattern, where two of the class are known to be realized in the literature by the underlying family symmetries of the $D_4$ and $A_4$ groups, the dihedral and tetrahedral symmetry groups..

**Abstrak**
Mengacu pada hasil-hasil eksperimen terkini, seperti osilasi neutrino dan peluruhan beta ganda tanpa neutrino (misalnya data dari Super-Kamiokande, KamLAND, SNO, dsb.), matriks campuran neutrino telah dapat ditentukan. Walaupun masih ada beberapa parameter tertentu yang mempunyai nila-nilai batas yang lebar, tetapi berdasarkan hasil eksperimen terkini, dimungkinkan untuk menyusun sebuah bentuk umum matrik massa neutrino.
Berawal dari bentuk umum matrik massa neutrino ini, dengan menerapkan beberapa syarat tertentu dalam konteks mekanisme seesaw ditentukan kemungkinan pola-pola matrik massa neutrino yang memiliki tekstur nol. Dari pola matrik massa neutrino yang diperoleh, terdapat tiga kelas pola, yang mana dua dari kelas tersebut diketahui terealisasi dalam literatur-literatur dengan simetri keluarga penopangnya adalah grup $D_4$ dan $A_4$, yaitu grup simetri dihedral dan tetrahedral.

**Kata kunci**: matriks massa neutrino, mekanisme seesaw, simetri keluarga.

## 1. Introduction

According to the Standard Model (SM) of Particle Physics, based on gauge symmetry $SU(3)_C \otimes SU(2)_L \otimes U(1)_Y$ neutrinos are massless, neutral, and only left-handed neutrinos participate in weak and electromagnetic interactions via charge-current and neutral-current interactions. Even though the SM is very successful in both describing wide range of phenomena and its predictions on the related physical quantities, there is no fundamental reason to put neutrino to be massless like photon to be massless due to $U(1)_{EM}$ gauge invariance.

For more than two decades the solar neutrino flux measured on Earth has been much less than that predicted by the solar model[1]. From recent experimental results, there is evidence that neutrinos are massive due to the experimental facts that the solar and atmospheric neutrinos undergo oscillations during its propagation[2,3,4]. Theoretically, neutrino oscillations can be explained by using quantum mechanics and occurs if non-zero $\Delta m_{ij}^2$, where $\Delta m_{ij}^2 = m_i^2 - m_j^2$ ($i, j = 1, 2, 3$, $m_i$ and $m_j$ are the neutrino mass in the mass eigenstates basis) and the neutrino flavor eigenstates are different from neutrino mass eigenstates. The flavor eigenstates $(\nu_e, \nu_\mu, \nu_\tau)$ can be expressed as a linear combination of neutrino mass eigenstates $(\nu_1, \nu_2, \nu_3)$ as follows:

$$\begin{pmatrix} \nu_e \\ \nu_\mu \\ \nu_\tau \end{pmatrix} = V \begin{pmatrix} \nu_1 \\ \nu_2 \\ \nu_3 \end{pmatrix}, \qquad (1)$$

where $V$ is a neutrino mixing matrix like the *Cabibbo-Kobayashi-Maskawa* (CKM ) matrix in the quark sector.

From the theoretical side, several models have been proposed to explain the existence of the neutrino mass and its underlying symmetries. Among the proposed models, the one based on seesaw mechanism is the most popular because it can explain the smallness of the neutrino mass and also account for the large mixing angles simultaneously[5].

Recently, one of the interesting subjects is the determination of the lepton family symmetries underlying the neutrino mass patterns in the seesaw mechanism. The general neutrino mass matrix $M_\nu$ in the seesaw mechanism is given by[6]:

$$M_\nu = M_D M_N^{-1} M_D^T, \qquad (2)$$

where $M_D$ and $M_N$ are the Dirac mass and Majorana mass matrices respectively. Based only on phenomenological consideration, Ma proposed the *All-purpose neutrino mass matrix* [7] and then he also derived the same neutrino mass matrix by using the cubic symmetry group $S_4$ (contain $A_4$ and $S_3$ groups) as the family symmetry[8]. Several models based on the tetrahedral group $A_4$ has been proposed also by Zee[9].

In this paper, we derive the general neutrino mass matrix patterns from the neutrino mixing matrix $V$, suitable with the current experimental limits (Section 2). In Section 3, using the general neutrino mass matrix patterns obtained in section 2, we determine the possible structure of the $M_N^{-1}$ matrix from the general neutrino mass matrix pattern obtained in section 2, but with additional requirement that is $M_N$ has one or more texture zero. We mention also the corresponding underlying family symmetry for the neutrino mass matrix. Finally, in Section 4 we give the conclusion.

## 2. General Neutrino Mass Matrix Patterns

Following standard convention, let us denote the neutrino current eigenstates coupled to the charge leptons by the $W$ bosons as $\nu_\alpha$ $(\alpha = e, \mu, \tau)$, and the neutrino mass eigenstates as $\nu_i$ $(i = 1, 2, 3)$ then the mixing matrix $V$ takes the form presented in Eq. (1). As we have stated explicitly above, we use the seesaw mechanism as the responsible mechanism for generating the neutrinos mass. The Majorana-mass Lagrangian term is given by:

$$L = -\nu_\alpha M_{\alpha\beta} C \nu_\beta + h.c. \qquad (3)$$

where $C$ denotes the charge conjugation matrix. Thus, the neutrino mass matrix $M$ is symmetric. For the sake of simplicity we will assume *CP* conservation so that $M$ is real. Within this simplification, the neutrino mass matrix $M$ is diagonalized by the orthogonal transformation:

$$V^T M V = \begin{pmatrix} m_1 & 0 & 0 \\ 0 & m_2 & 0 \\ 0 & 0 & m_3 \end{pmatrix}. \qquad (4)$$

From recent experimental results, the explicit form of the mixing matrix moduli $V$ has been known to be[10]:

$$|V| = \begin{pmatrix} 0.79 - 0.88 & 0.47 - 0.61 & < 0.20 \\ 0.19 - 0.52 & 0.42 - 0.73 & 0.58 - 0.82 \\ 0.20 - 0.53 & 0.44 - 0.74 & 0.56 - 0.81 \end{pmatrix}. \qquad (5)$$

According to the requirement that the mixing matrix $V$ must be orthogonal, together with putting $V$ in nice simple looking numbers, the following form of $V$ are used to be proposed,

$$V = \begin{pmatrix} -2/\sqrt{6} & 1/\sqrt{3} & 0 \\ 1/\sqrt{6} & 1/\sqrt{3} & 1/\sqrt{2} \\ 1/\sqrt{6} & 1/\sqrt{3} & -1/\sqrt{2} \end{pmatrix}, \qquad (6)$$

we will use this form of $V$ as our mixing matrix. To determine the neutrino mass matrix $M_\nu$, we employ the following relation:

$$M_\nu = V \left( V^T M V \right) V^T. \qquad (7)$$

Substitutions of Eq. (4) and Eq. (6) into Eq. (7) lead to:

$$M_\nu = \begin{pmatrix} -2/\sqrt{6} & 1/\sqrt{3} & 0 \\ 1/\sqrt{6} & 1/\sqrt{3} & 1/\sqrt{2} \\ 1/\sqrt{6} & 1/\sqrt{3} & -1/\sqrt{2} \end{pmatrix} \begin{pmatrix} m_1 & 0 & 0 \\ 0 & m_2 & 0 \\ 0 & 0 & m_3 \end{pmatrix} \begin{pmatrix} -2/\sqrt{6} & 1/\sqrt{6} & 1/\sqrt{6} \\ 1/\sqrt{3} & 1/\sqrt{3} & 1/\sqrt{3} \\ 0 & 1/\sqrt{2} & -1/\sqrt{2} \end{pmatrix} \tag{8}$$

$$= \begin{pmatrix} \dfrac{2m_1+m_2}{3} & \dfrac{m_2-m_1}{3} & \dfrac{m_2-m_1}{3} \\ \dfrac{m_2-m_1}{3} & \dfrac{m_1+2m_2+3m_3}{6} & \dfrac{m_1+2m_2-3m_3}{6} \\ \dfrac{m_2-m_1}{3} & \dfrac{m_1+2m_2-3m_3}{6} & \dfrac{m_1+2m_2+3m_3}{6} \end{pmatrix}.$$

Inspecting Eq. (8), one can write the general neutrino mass matrix patterns as follows:

$$M_\nu = \begin{pmatrix} a & b & b \\ b & c & d \\ b & d & c \end{pmatrix}. \tag{9}$$

From Eq. (9) we can see that the general neutrino mass matrix obtained is symmetric as required.

## 3. Possible Family Symmetries of the General Neutrino Mass Matrix

Having formulated the general neutrino mass matrix, we proceed to determine the kinds of the possible underlying family symmetries of the general neutrino mass matrix appearing in Eq. (9). In the seesaw mechanism scheme, the neutrino mass matrix is given by Eq. (2). If the general neutrino mass matrix in Eq. (9) is taken as the neutrino mass matrix generated by seesaw mechanism, then we can write $M_\nu$ as follows:

$$M_\nu = \begin{pmatrix} a & b & b \\ b & c & d \\ b & d & c \end{pmatrix} = M_D M_N^{-1} M_D^T. \tag{10}$$

The kinds of the family symmetries can be determined from the structure of the Majorana mass matrix $M_N$ as well as the Dirac mass matrix $M_D$. We assume $M_N$ has one or more of its element to be zero (texture zero). The textures zeros of the mass matrix indicate the existence of additional symmetries beyond the SM[11]. If $M_N$ matrix has one or more of its element to be zero (textures zeros), then this implies $M_N^{-1}$ matrix has one or more $2\times2$ sub-matrices with zero determinants.

If we take $M_D$ to be diagonal:

$$M_D = \begin{pmatrix} u_1 & 0 & 0 \\ 0 & u_2 & 0 \\ 0 & 0 & u_3 \end{pmatrix},$$

with the condition $u_2 = u_3 = u$ (CP is conserved in neutrino oscillations)[12], then the property of $M_N^{-1}$ matrix is preserved in $M_\nu$[13]. According to the above requirements, we have the $M_N^{-1}$ matrix pattern as:

$$M_N^{-1} = \begin{pmatrix} A & B & B \\ B & C & D \\ B & D & C \end{pmatrix}. \tag{11}$$

Now we put the texture zero requirements on the $M_N$ The possible patterns of $M_N$ matrix from Eq. (11), is:

$$M_N = \frac{1}{A(C^2-D^2)+2B^2(D-C)} \begin{pmatrix} C^2-D^2 & B(D-C) & B(D-C) \\ B(D-C) & AC-B^2 & -AD+B^2 \\ B(D-C) & -AD+B^2 & AC-B^2 \end{pmatrix} \tag{12}$$

The $M_N$ matrix in Eq. (12) will have textures zero if one or more of the following relations are satisfied:

i. $C = -D$,

ii. $AD - B^2 = 0$,

iii. $AC - B^2 = 0$,

iv. $B = 0$.

$$\tag{13}$$

Once we can determine the form of $M_N$ matrix, then the corresponding $M_N^{-1}$ and $M_\nu$ matrices can be obtained.

**Case** (i) $C = -D$, we obtain:

$$M_N = \begin{pmatrix} 0 & \dfrac{1}{2B} & \dfrac{1}{2B} \\ \dfrac{1}{2B} & \dfrac{-AD-B^2}{4B^2D} & \dfrac{-AD+B^2}{4B^2D} \\ \dfrac{1}{2B} & \dfrac{-AD+B^2}{4B^2D} & \dfrac{-AD-B^2}{4B^2D} \end{pmatrix} \quad (14)$$

The $M_N$ matrix in Eq. (14) has the corresponding $M_N^{-1}$ and $M_\nu$ matrices as follows:

$$M_N^{-1} = \begin{pmatrix} A & B & B \\ B & -D & D \\ B & D & -D \end{pmatrix}, \quad (15)$$

and

$$M_\nu = \begin{pmatrix} P & Q & Q \\ Q & -R & S \\ Q & S & -R \end{pmatrix} \quad (16)$$

where $P = A/u_i^2$, $Q = B/u_1 u$, and $R = S = D/u^2$.

**Case** (ii) $AD = B^2$, we obtain:

$$M_N = \begin{pmatrix} \dfrac{C+D}{A(C-D)} & \dfrac{-B}{A(C-D)} & \dfrac{-B}{A(C-D)} \\ \dfrac{-B}{A(C-D)} & \dfrac{1}{C-D} & 0 \\ \dfrac{-B}{A(C-D)} & 0 & \dfrac{1}{C-D} \end{pmatrix} \quad (17)$$

with

$$M_N^{-1} = \begin{pmatrix} A & B & B \\ B & C & D \\ B & D & C \end{pmatrix} \quad (18)$$

and

$$M_\nu = \begin{pmatrix} P & Q & Q \\ Q & R & S \\ Q & S & R \end{pmatrix} \quad (19)$$

where $P = A/u_i^2$, $Q = B/u_1 u$, $R = C/u^2$, and $S = D/u^2$.

**Case** (iii) $AC = B^2$, we obtain:

$$M_N = \begin{pmatrix} \dfrac{C+D}{A(C-D)} & \dfrac{-B}{A(C-D)} & \dfrac{-B}{A(C-D)} \\ \dfrac{-B}{A(C-D)} & 0 & \dfrac{-1}{C-D} \\ \dfrac{-B}{A(C-D)} & \dfrac{-1}{C-D} & 0 \end{pmatrix} \quad (20)$$

with

$$M_N^{-1} = \begin{pmatrix} A & B & B \\ B & C & D \\ B & D & C \end{pmatrix} \quad (21)$$

and

$$M_\nu = \begin{pmatrix} P & Q & Q \\ Q & R & S \\ Q & S & R \end{pmatrix} \quad (22)$$

where $P = A/u_i^2$, $Q = B/u_1 u$, $R = C/u^2$, and $S = D/u^2$.

**Case** (iv) $B = 0$, we obtain:

$$M_N = \begin{pmatrix} 1/A & 0 & 0 \\ 0 & \dfrac{C}{C^2-D^2} & \dfrac{-D}{C^2-D^2} \\ 0 & \dfrac{-D}{C^2-D^2} & \dfrac{C}{C^2-D^2} \end{pmatrix} \quad (23)$$

with

$$M_N^{-1} = \begin{pmatrix} A & 0 & 0 \\ 0 & C & D \\ 0 & D & C \end{pmatrix} \quad (24)$$

and

$$M_\nu = \begin{pmatrix} P & 0 & 0 \\ 0 & R & S \\ 0 & S & R \end{pmatrix} \quad (25)$$

where $P = A/u_i^2$, $R = C/u^2$, and $S = D/u^2$.

**Case** (v) $C = -D$, and $AD = B^2$, we obtain:

$$M_N = \begin{pmatrix} 0 & 1/2B & 1/2B \\ 1/2B & -1/2D & 0 \\ 1/2B & 0 & -1/2D \end{pmatrix} \quad (26)$$

with

$$M_N^{-1} = \begin{pmatrix} A & B & B \\ B & -D & D \\ B & D & -D \end{pmatrix} \quad (27)$$

and

$$M_\nu = \begin{pmatrix} P & Q & Q \\ Q & -R & R \\ Q & R & -R \end{pmatrix} \quad (28)$$

where $P = A/u_i^2$, $Q = B/u^2$, and $R = D/u^2$.

**Case** (vi) $C = -D$, and $AC = B^2$, we obtain:

$$M_N = \begin{pmatrix} 0 & 1/2B & 1/2B \\ 1/2B & 0 & -1/2C \\ 1/2B & -1/2C & 0 \end{pmatrix} \quad (29)$$

with

$$M_N^{-1} = \begin{pmatrix} A & B & B \\ B & C & -C \\ B & -C & C \end{pmatrix} \quad (30)$$

and

$$M_\nu = \begin{pmatrix} P & Q & Q \\ Q & R & -R \\ Q & -R & R \end{pmatrix} \quad (31)$$

where $P = A/u_1^2$ , $Q = B/u^2$ , and $R = D/u^2$ .

**Case** $(\text{vii})$ $B = 0$, and $AD = B^2$, we obtain:

$$M_N = \begin{pmatrix} 1/A & 0 & 0 \\ 0 & 1/C & 0 \\ 0 & 0 & 1/C \end{pmatrix} \qquad (32)$$

with

$$M_N^{-1} = \begin{pmatrix} A & 0 & 0 \\ 0 & C & 0 \\ 0 & 0 & C \end{pmatrix} \qquad (33)$$

and

$$M_\nu = \begin{pmatrix} P & 0 & 0 \\ 0 & R & 0 \\ 0 & 0 & R \end{pmatrix} \qquad (34)$$

where $P = A/u_1^2$ and $R = C/u^2$ .

**Case** $(\text{viii})$ $B = 0$, and $AC = B^2$, we obtain:

$$M_N = \begin{pmatrix} 1/A & 0 & 0 \\ 0 & 0 & 1/D \\ 0 & 1/D & 0 \end{pmatrix} \qquad (35)$$

with

$$M_N^{-1} = \begin{pmatrix} A & 0 & 0 \\ 0 & 0 & D \\ 0 & D & 0 \end{pmatrix} \qquad (36)$$

and

$$M_\nu = \begin{pmatrix} P & 0 & 0 \\ 0 & 0 & S \\ 0 & S & 0 \end{pmatrix} \qquad (37)$$

where $P = A/u_1^2$ , and $S = D/u^2$ .

From the eight cases considered above, we can see that the structure of the $M_N^{-1}$ and $M_\nu$ matrices could be classified into three classes due to their textures whether it contains textures zeros or not, they are:

1. Both $M_N^{-1}$ and $M_\nu$ matrices have no textures zeros, it appears in the cases: $C = -D$, $AD = B^2$, $AC = B^2$, $C = -D$ combined with $AD = B^2$, and $C = -D$ combined with $AC = B^2$ .

2. Both $M_N^{-1}$ and $M_\nu$ matrices have four zeros as can be seen in Eqs. (24) and (25) for the case $B = 0$ .

3. Both $M_N^{-1}$ and $M_\nu$ matrices have six zeros as can be seen in Eqs. (33) and (34) for the case $B = 0$ combined with $AD = B^2$, and in Eqs. (36) and (37) for the case $B = 0$ combined with $AC = B^2$ .

Specifying the underlying family symmetry required specifying the particle contents of a model and the corresponding representation for each particle. There are numerous papers which start from certain family symmetry group in the quark and lepton sectors or only in the lepton sectors that tried to derive the mixing matrix and the neutrino mass matrix from the family symmetry groups. We can mention some of them, for example E. Ma[14] gives a models based on A$_4$ where he puts the lepton family doublet $(\nu_i, l_i)$, and the singlet $l_i^c$ ( $i = 1, 2, 3$ ) to be $\underline{3}$ and additional three heavy neutral fermion singlet $N_i$ to be $\underline{1}$ , $\underline{1}$', $\underline{1}$'', this model gives $M_N$ in the form of Eq. (35) correspond to the case (viii). In other paper, he used the same group A$_4$ but assigning $(\nu_i, l_i)$ to be $\underline{3}$ and $l_i^c$ to be $\underline{1}$', $\underline{1}$'', together with three Higgs doublet to be $\underline{3}$ and one Higgs singlet to be $\underline{1}$, he obtain a similar pattern.

The dihedral group D$_4$ has been used also as the underlying family symmetry, in one of the model by Ma[14]. In this model $(\nu_i, l_i)$ and $\nu_i^c$ are assigned to the $\underline{1}^{++} + \underline{2}$ representation of D$_4$ group. The Higgs sector has 3 doublets: $\Phi_{1,2,3} \sim \underline{1}^{++}, \underline{1}^{++}, \underline{1}^{+-}$ and 2 singlets: $\chi_{1,2} \sim \underline{2}$ . Adding an extra Z$_2$ symmetry such that the $l_1^c, \nu_{1,2,3}^c, \Phi_1$ are odd, while all other fields are even, the charge lepton mass matrix is diagonal with 3 independent eigenvalues and the neutrino mass matrix was similar to case (vii).

Those two groups above are just some of the underlying family symmetry group that can realize the structure of the $M_\nu$ matrices for both class 1 and 3. Meanwhile, the underlying family symmetry to be responsible for $M_N^{-1}$ and $M_\nu$ matrices structure in class 2 have not been considered in any paper that we know of. This may be a subject for further research.

## 4. Conclusion

By choosing a nice looking form of the mixing matrix elements $V$, fulfilling orthogonality condition, together with the condition that the Dirac mass matrix is diagonal with degenerate $u_2 = u_3 = u$ , we obtain the general neutrino mass matrix pattern $M_\nu$ .

Assuming that the resulting general neutrino mass matrix to be a mass matrix generated by a seesaw mechanism combined with the requirements that the Majorana mass matrix has one or more textures zeros, we find that there are eight possible cases for the $M_N$ matrix. Based on the structure of the $M_N^{-1}$ matrices whether it contains textures zeros or not, the eight possible patterns of the $M_N$ matrix lead to the three classes of the $M_\nu$ matrix.

Two of the $M_\nu$ classes (class 1 and 3) are known to have the $A_4$ and $D_4$ groups to be their possible underlying family symmetries. While the underlying family symmetry that can realize the $M_\nu$ matrix in class 2 were not known in any literature.

**Acknowledgment**

The first author would like to thank the Physics Department of the Gadjah Mada University where he is currently a graduate student.

**References**

1.  Pantaleone, J., *Three Neutrino Flavors: Oscillations, Mixing, and the Solar Neutrino Problem,* Phys. Rev. **D 43**, R641-R645, 1991.
2.  Fukuda, Y., *et.al, Measurements of the Solar Neutrino Flux from Super-Kamiokande's First 300 Days*, Phys. Rev. Lett. **81**, 1158-1562, 1998.
3.  Fukuda, Y., *et.al, Measurement of the Solar Neutrino Energy Spectrum Using Neutrino-Electron Scattering*, Phys. Rev. Lett. **82**, 2430-2434, 1999.
4.  Ahn, M.H., *et.al, Indications of Neutrino Oscillation in a 250* km *Long-Baseline Experiment,* Phys. Rev. Lett. **90**, 041801 (1-5), 2003.
5.  Tsujimoto, H., *Absolute Values of Neutrino Masses Implied by the Seesaw Mechanism,* hep-ph/0501023.
6.  Altarelli, G., and Feruglio, F., *Models of Neutrino Masses and Mixings*, hep-ph/0405048.
7.  Ma, E., *All-purpose neutrino mass matrix,* Phys. Rev. **D 66**, 117301(1-3), 2002.
8.  Ma, E., *Neutrino Mass Matrix from $S_4$ Symmetry*, hep-ph/0508231.
9.  Zee, A., *Obtaining the Neutrino Mixing Matrix with the Tetrahedral Group*, hep-ph/0508278.
10. Gonzales-Garcia, M. C., *Global Analysis of Neutrino Data,* hep-ph/0410030.
11. Leontaris, G. K., Lola, S., Scheich, C., and Vergados, J. D., *Textures for Neutrino Mass Matrices*, Phys. Rev. **D53**, 6381-6397, 1996.
12. Ma, E., *Tetrahedral Family Symmetry and the Neutrino Mixing Matrix,* hep-ph/0508099.
13. Ma, E., *Connection Between the Neutrino Seesaw Mechanism and Properties of the Majorana Neutrino Mass Matrix*, Phys. Rev. **D71**, 2005.
14. Ma, E., *Non-Abelian Discrete Family Symmetries of Leptons and Quarks*, hep-ph/0409075.